\documentclass[aps,prb,twocolumn,floatfix,amssymb]{revtex4}

\usepackage{amsmath}
\usepackage{amsxtra}
\usepackage{graphicx}

\begin{document}

\title{Spin torque contribution to the ac spin Hall conductivity}   
\author{Arturo Wong{\email{wong@ccmc.unam.mx}}, Jes\'us A. Maytorena, Catalina L\'opez-Bastidas and Francisco Mireles}     
           
\affiliation{Departamento de F{\'\i}sica Te\'orica, Centro 
de Ciencias de la Materia Condensada, Universidad Nacional 
Aut\'onoma de M\'exico, Apdo. Postal 2681, 22800 Ensenada, 
Baja California, M\'exico}

\date{\today}          

\begin{abstract}
Using the recently proposed definition of a conserved spin-current operator 
[J. Shi {\it et al.}, Phys. Rev. Lett. {\bf 96}, 076604 (2006)] we explore the frequency dependent spin Hall conductivity for a two-dimensional electron 
gas with Rashba and Dresselhaus spin-orbit interaction in response to an oscillating electric field. We show  that the optical spectrum of the spin Hall conductivity exhibit  remarkable changes  when the new definition of spin current is applied. Such behavior is mainly due  to a significant contribution of the spin torque term which is absent in the conventional form of the spin current. In addition, it is observed that the magnitude and direction of the dynamic spin Hall current strongly depends on the electric field frequency as with the interplay of the spin-orbit coupling strengths. 
\end{abstract}

\maketitle

\section{INTRODUCTION}  

The spin-Hall effect (SHE) is a phenomenon that has and still motivating a very conspicuous interest among the semiconductor spintronics\cite{Wolf,Zutic} researchers. The SHE refers to a spin-accumulation induced via a driven spin current in response to a  perpendicular dc electric field in finite width nonmagnetic systems which experiment spin-orbit interaction (SOI) effects.\cite{Hirsch,S.Murakami,J.Sinova} This spin (Hall) accumulation with opposite spin-magnetization at the edges of such systems has been beautifully demonstrated by recent optical experiments.\cite{Wunderlich,Kato} These observations were followed by purely electrical measurements of the SHE in metallic conductors, \cite{Valenzuela} and by its very recent detection at room temperature via Kerr spectroscopy in  {\it n}-type ZnSe samples.\cite{Stern} 

The role of disorder in the universal characteristic\cite{Sinova1} of the  dc spin-Hall conductivity ($e/8\pi$) in two dimensional electron gases (2DEG's) with Rashba SOI has been a central topic. Nowadays there is a general consensus that the spin-Hall conductivity is suppressed only in the static limit even in arbitrary weak disordered systems and as long as $k$-linear Rashba SOI is considered in infinite size samples.\cite {Sinova1,Inoe1,Mishchenko,Chalaev-Loss,Dimitrova,Raimondi} 
This however does not hold in the presence of magnetic fields and/or magnetic impurities,\cite{Sinova1,Inoe2,Pei-Wang} neither for 2D hole systems\cite{Schliemann,Bernevig,Khaetskii} for which the static spin-Hall conductivity is robust against disorder. Recently Grimaldi {\it et al.}\cite{Grimaldi1} have showed that this is also true when the  Fermi energy is comparable to the spin-orbit splitting energy ($E_F\sim \Delta$).

Yet another possibility to gain further physical insight to the SHE is the finite frequency regime of transport.\cite{Sherman} 
Recent studies of the ac\,\,field-induced charge \cite{Xu} and spin\cite{Wang} response of 2DEG's have emphasized the importance of the dynamic regime.\cite{Maytorena-etal} It has been suggested that an ac\,\,probing field  can be used to control the spin-Hall current in 2DEG's with Rashba and/or Dresselhaus SOI.\cite{Wang,Maytorena-etal} It turns out that for finite frequencies (in the THz range), the cancellation of the intrinsic spin Hall effect due to impurity scattering \cite{Khaestkii1} is no longer perfect, and in principle the effect may survive.\cite{Finkelstein}

Sugimoto {\it et al.} \cite{Sugimoto}, on the other hand, have raised the question recently about the dependence of the vanishing of the spin-Hall conductivity $\sigma_{sH}(0)$ on the actual definition of the spin-current operator used. This is certainly a fundamental issue that we extend here at the finite frequency regime of the spin-Hall conductivity, $\sigma_{sH}(\omega)$. The widely {\it ad hoc} form  used for the spin current operator for an electronic system 
J$_{s}=\frac{\hbar}{4} \{ \sigma_z ,v_y\}$, where $\sigma_z$ is the Pauli spin $z-$component and  $v_y$ the electron velocity operator, respectively, has  the desirable form that resembles the charge-current operator. In 2D holes gases the conventional definition seems to be consistent with edge-spin accumulation experiments\cite{Nomura} as with measurements of optically injected spin-polarized currents in semiconductors. \cite{Stevens-Sipe}
However, it does not fulfill a simple continuity equation\cite{Murakami2,Culcer,Rashba2} and appears as incomplete in describing spin transport in systems with spin-orbit coupling.\cite{Shi}

%However, it turns out to be unphysical since it is ill-defined in the presence of SOI effects, leading to a concomitant failure of the continuity equation for the spin density.\cite{Murakami2,Culcer,Rashba2}

Many efforts to clarify this issue have gone into this direction lately.\cite{Shi,Shen,Sun,Jin,Li,Yang,Chen,Zhang,Wang1} 
It seems nevertheless  an unsettled problem yet. Furthermore, the discussions have been focused namely in the zero frequency limit. In particular, the recent proposal by J. Shi {\it et al.} \cite{Shi} of a dubbed unambiguous and proper definition of a ``conserved" spin current density, is indeed physically appealing and certainly deserves a close examination. Such new definition adds to the conventional part {\bf J}$_s$, a spin source term (torque dipole density {\bf P}$_{\tau}$) associated to the electron spin precessional motion with the total spin current density defined as ${\cal J}_s=$ {\bf J}$_s+${\bf P}$_{\tau}$. Moreover, among other interesting properties, it ensures that a continuity equation $\frac{\partial S_z}{\partial t}+\nabla\cdot{\cal J}_s=0  $ is always satisfied, with  $S_z=\frac{\hbar}{2}\Psi^\dagger\sigma_z\Psi$ describing the spin density. It is thus worth to elucidate to what extent the use of this conserved spin current operator provides new insight on the ac spin-Hall conductivity response in 2D systems with SOI effects. Whether or not such new definition of the spin current operator offers a physically satisfactory description of the ac spin Hall conductivity should be ultimately validated by the experiment.

In this paper we study the ac\,\,spin-Hall conductivity within the linear response theory employing the new definition for the spin current operator reported by J. Shi et al.\cite{Shi} for a 2DEG in the presence of Rashba and Dresselhaus SOI and subject to an  alternating electric field of frequency $\omega$. It is verified {\it post facto}  that the optical spectrum of the spin Hall conductivity changes drastically when the new definition is used due to a non-trivial spin-torque contribution and to the interplay of the Rashba and Dresselhaus SOI. We predict that the torque dipole contribution to the ac spin-Hall conductivity generally dominates over the conventional part in typical samples.  

The work is organized as follows. In Sec. II we describe the Hamiltonian model for a 2DEG in the presence of Rashba and Dresselhaus spin-orbit interaction. The main features of the Kubo formula in linear response applied to the spin-Hall conductivity are presented in Sec. III. An analysis of the calculated ac spin-Hall response using the conserved spin-current operator is presented also in this section. In Sec. IV we give a discussion of the analytical and numerical results obtained for typical 2DEG systems. A 
Summary is given in Sec. V and we end with a brief Appendix which outline the derivation of the spin-current-charge-current correlation function used in the Kubo formula.

\section{THE MODEL}

We consider a two-dimensional (2D) electron system with a single-particle 
Hamiltonian given by

\begin{equation}
\label{Full-Hamiltonian}
H=\frac{\ p^{2}}{\ 2 m^*}+H_{so}
\end{equation}

\noindent where the spin-orbit part $H_{so}$ is the sum of the Rashba and 
(linear) Dresselhaus SOI type with

\begin{equation}
H_{so}=\frac{\ \alpha}{\ \hbar} (\sigma_y p_x-\sigma_x p_y)+\frac{\ \beta}{\ \hbar} (\sigma_x p_x-\sigma_y p_y)
\end {equation}

\begin{figure}  
\centerline{\includegraphics[width=3 in]{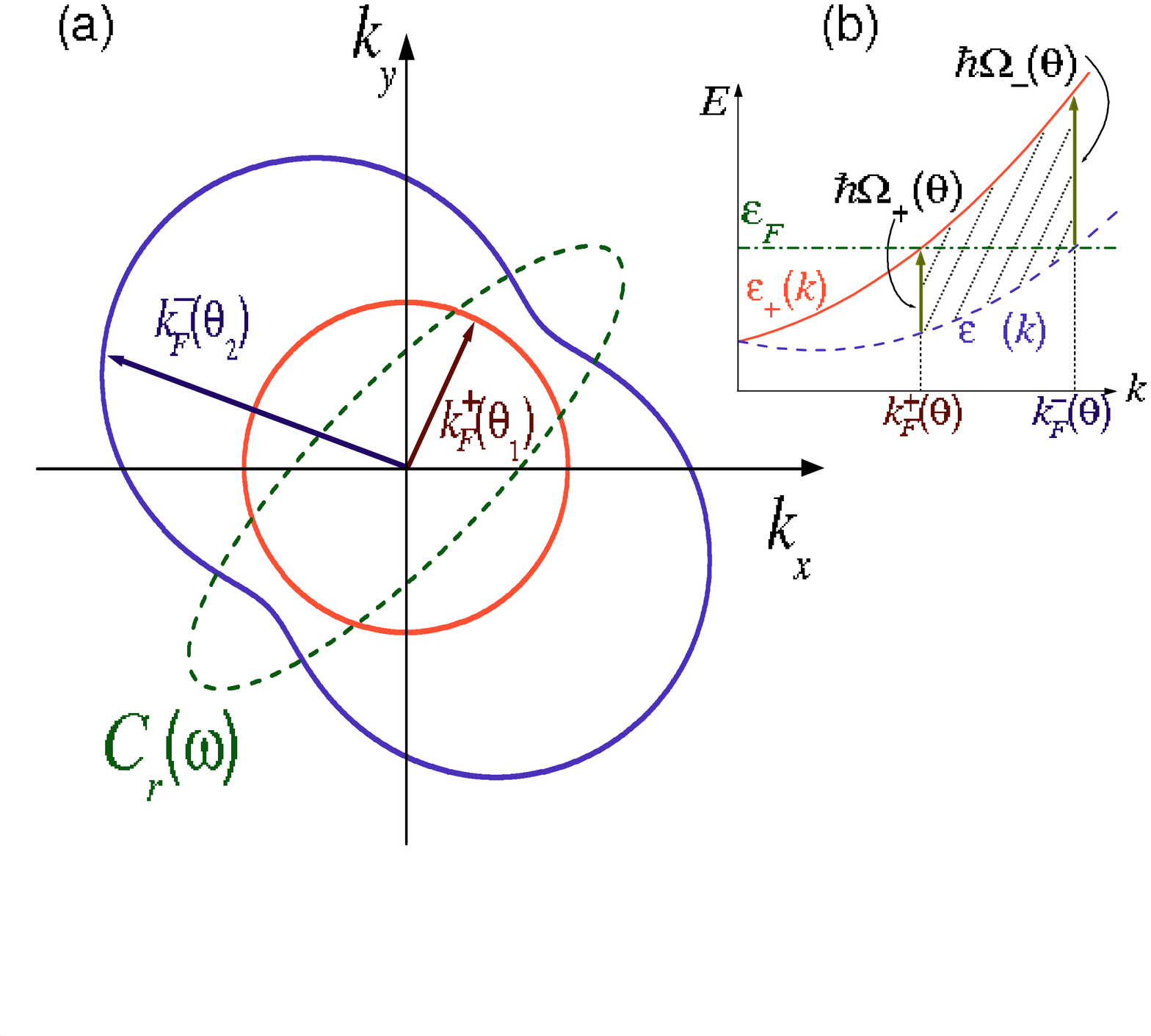}}
\caption{(Color online) (a) Schematic diagram showing the angular anisotropy of 
the spin-split conduction bands  at the Fermi energy in a 2DEG with finite 
Rashba and Dresselhaus spin-orbit interaction. The dotted $C_r(\omega)$ curve is a rotated ellipse resulting from the condition $\varepsilon_+(${\bf k}$)-\varepsilon_-(${\bf k}$)=\hbar\omega$, see text for details.  (b) Diagram depicting the allowed optical transitions (shadowed region) between the initial $\varepsilon_-$ and final $\varepsilon_+$ spin-split branches at a fixed polar angle $\theta$.}
\label{Fig1}
\end{figure}

\noindent here, $\alpha$ and $\beta$ are the coupling strengths for the 
Rashba and Dresselhaus couplings, respectively, with $p_x$, $p_y$ the 
components of the 2D momentum operator and $\sigma_x$, $\sigma_y$ the 
usual spin Pauli matrices.  For the full Hamiltonian (\ref{Full-Hamiltonian}) 
the energy spectrum  is simply 

\begin{equation}
\varepsilon_\mu=\frac{\ \hbar^2 k^2}{\ 2m^*}+\mu k \Delta (\theta)
\end{equation}

\noindent with $\mu \in\{+1,-1\}$ denoting the spin-states for spin-split 
dispersion bands and 
$\Delta (\theta)=\sqrt{(\alpha^2+\beta^2)\lambda(\theta)} $
describes the angular anisotropy of the spectrum (see Fig. 1) with  

\begin{equation}
\label{lambda}
\lambda(\theta)=1-\frac{\ 2\alpha \beta}{\ \alpha^2+\beta^2} { \sin } 2\theta
\end{equation}

\noindent and $\theta =$ tan$^{-1}\left( \frac{\ k_y}{\ k_x}\right)$. 
When $\alpha$ or $\beta$ is null, $\lambda(\theta)=1$ and the spin
splitting in $k-$space becomes isotropic. 
The eigenstates have form
 
\begin{equation}
\label{eigenvectors}
\vert \psi_{{\bf k},\mu}(\textbf{r})\rangle=\frac{\ e^{i{\bf k} 
\cdot {\bf r}}}{\ \sqrt{2 A}} 
\begin{bmatrix} 1 \\ 
-\mu e^{-i \phi} \\ \end{bmatrix} 
\end{equation}

\noindent with the wave vector {\bf k}$=(k_x,k_y)=k(\cos\theta,\sin\theta)$ in polar coordinates, $\phi=$ tan$^{-1}\left( \frac{\ \alpha k_x-\beta k_y}
{\ \alpha k_y-\beta k_x}\right) $ and $A$ the area of the system. 
At zero temperature, the two spin-split subbands are filled up to the 
same (positive) Fermi energy level $\varepsilon_F$ but with different 
Fermi wave vectors $k_F^{\mu}(\theta)= \sqrt{\frac{\ 2m^*\varepsilon_F}
{\ \hbar^2}+k_{so}^2 \lambda(\theta)}-\mu k_{so}\sqrt{\lambda(\theta)}$. 
Here, $\varepsilon_F = \frac{\ \hbar^2 k^2_F}{\ 2m^*}$ with 
$k_F=\sqrt{2\pi n-2k_{so}^2}$ and $n$ being the electron density. 
The SOI introduces a characteristic SOI energy and wavevector given by
$\varepsilon_{so} = \frac{\ \hbar^2 k^2_{so}}{\ 2m^*}$ and
$k_{so}=\frac{\ m^* \sqrt{\alpha^2+\beta^2}}{\ \hbar^2}$, respectively.

\section{Spin Hall conductivity at finite frequency}  

Within linear response to a weak (spatially homogeneous) electric field of 
frequency $\omega$ oscillating in the plane of the 2DEG, the frequency 
dependent spin-Hall conductivity can be calculated  using the standard Kubo 
formula\cite{Schliemann} in the limit $q\rightarrow 0$,

\begin{multline}\label{kubo}
\sigma_{xy}^{S,z}(\omega)=\frac{\ e}{\ \hbar A (\omega+i\eta)} 
\int^\infty_{0} e^{i(\omega+i\eta)t} \\
\times
\sum_{{\bf k},\mu}f(\varepsilon_\mu)_{T=0}\langle 
\psi_{{\bf k},\mu}(\textbf{r}) \vert [{\cal J}_x^{S,z}(t), v_y (0)] 
\vert \psi_{{\bf k},\mu}(\textbf{r}) \rangle dt ,
\end{multline}

%\begin{multline}\label{kubo}
%\sigma_{xy}^{S,z}(\omega)=\frac{\ e}{\ \hbar A (\omega+i\eta)} 
%\int^\infty_{0} e^{i(\omega+i\eta)t}\\ 
%\times\sum_{{\bf k},\mu}f(\varepsilon_\mu)_{T=0}\langle 
%\psi_{{\bf k},\mu}(\textbf{r}) \vert [{\cal J}_x^{S,z}(t), v_y (0)] 
%\vert \psi_{{\bf k},\mu}(\textbf{r}) \rangle dt ,
%\end{multline}

\noindent where we have assumed zero temperature and non-interacting carriers. The parameter $\eta>0$ is here to regularize the integral, but it can be viewed (phenomenologically) as a measure of the electron momentum dissipation effects due to impurity scattering events or any other many-body interactions\cite{Schliemann} with an overall time relaxation rate $\tau=\eta^{-1}$ and $f(\varepsilon_\mu)$ is the Fermi distribution function.  Here we employ the conserved spin-current operator of Shi {\it et al.}\cite{Shi} written in the Heisenberg picture, ${\cal J}_x^{S,z}(t)$.
Unlike the conventional definition J$_s$, the new  effective spin current ${\cal J}_s$ is (i) in conjugation with a mechanical or thermodynamical force ({\it i.e.} Osanger's reprocity relations are well established), (ii) it has the desirable property that it vanishes for localized orbitals which anticipates a zero SHE for insulators, and  (iii) predicts that the static value of spin Hall conductivity has an opposite sign to those reported using the conventional definition for Rashba and linear Dresselhaus SOI models.\cite{Shi,Chen} It is shown in Ref.[\onlinecite{Shi}], after imposing a nule torque dipole density outside the bulk, that the effective spin current operator ${\cal \hat J}_s$ can be simply recasted as ${\cal \hat J}_s=\frac{\hbar}{2}\frac{d (\hat r \sigma_z)}{d t}$. As a result, it adds an extra term which takes into account the spin torque contribution.\cite{Shi} 
In \eqref{kubo} $v_y (0)$ is the single-particle operator for the electron velocity which can be obtained from the Heisenberg equation of motion {\bf v}$(0)=\frac{i}{\hbar}[ H,$ {\bf r}$]$. For our system we get  $v_y(0)=\frac{\ p_y}{\ m^*}-\Bigr(\frac{ \alpha \sigma_x+\beta \sigma_y}{\ \hbar}\Bigr)$.
The spin conductivity $\sigma_{xy}^{S,z}(\omega)$ in  \eqref{kubo} describes a $z$-polarized spin-current flowing in the $x$-direction as a function of frequency $\omega$ in response to a weak electric field $E(\omega)$  chosen along the $y$-direction.

To obtain explicitly ${\cal J}_x^{S,z}(t)$ we  first take the $x$-component of the current operator with a spin moment polarized along the $z$ direction at time $t=0$ (via the Heisenberg equation of motion), yielding

\begin{multline}
\label{corrienteespin}
{\cal J}_x^{S,z}(0)=\frac{\ \hbar}{\ 2m^*} \sigma_z p_x+\Bigr\{\hat{x}, \frac{\ \beta}{\ 2\hbar} (\sigma_x p_y+\sigma_y p_x) \\
-\frac{\ \alpha}{\ 2\hbar} (\sigma_x p_x+\sigma_y p_y)\Bigr\}
\end{multline}

%\begin{multline}\label{corrienteespin}
%{\cal J}_x^{S,z}(0)=\frac{\ \hbar}{\ 2m^*} \sigma_z p_x+\Bigr\{\hat{x}, \frac{\ %\beta}{\ 2\hbar} (\sigma_x p_y+\sigma_y p_x)\\
%-\frac{\ \alpha}{\ 2\hbar} (\sigma_x p_x+\sigma_y p_y)\Bigr\}
%\end{multline}

\noindent where the symbol $\{,\}$ denotes the anticommutator.  
The first term in \eqref{corrienteespin} represent the conventional 
spin-current part whiles the second term  arises entirely from the spin 
torque contribution.  

It can be shown that the expected value of the spin-current-charge-current 
correlation function in \eqref{kubo} $\langle \psi_{{\bf k},\mu}
(\textbf{r}) \vert [{\cal J}_x^{S,z}(t), v_y (0)] \vert \psi_{{\bf k},\mu}
(\textbf{r}) \rangle\equiv{\cal F}_{\mu}({\bf k},t)$ takes the form 
(see Appendix)

\begin{multline}
\label{valoresperado}
{\cal F}_{\mu}({\bf k},t)=i\mu \frac{\hbar k_x^2 
\left (\beta^2-\alpha^2\right )}{m^*k\Delta(\theta)} \Bigr [ \cos\Bigr(
\frac{\ 2k\Delta(\theta)}{\ \hbar}t\Bigr) \\
- \frac{\ 2k\Delta(\theta)}{\ \hbar}\,t\sin\Bigr(
\frac{\ 2k\Delta(\theta)}{\ \hbar}t\Bigr)\Bigr ]
\end{multline}

%\begin{multline}
%\label{valoresperado}
%{\cal F}_{\mu}({\bf k},t)=i\mu \frac{\hbar k_x^2 
%\left (\beta^2-\alpha^2\right )}{m^*k\Delta(\theta)} \Bigr [ \cos\Bigr(
%\frac{\ 2k\Delta(\theta)}{\ \hbar}t\Bigr) \\
%- \frac{\ 2k\Delta(\theta)}{\ \hbar}\,t\sin\Bigr(
%\frac{\ 2k\Delta(\theta)}{\ \hbar}t\Bigr)\Bigr ]
%\end{multline}

By using this result in \eqref{kubo} together with the fact that 
$\sum_{k,\mu}\mu f(\varepsilon_\mu)_{T=0}=-\sum_{k}
\Theta[(k-k_F^+)(k_F^--k)]$, where $\Theta(x)$ is the Heaviside step 
function, it is then possible to decompose the 
frequency dependent spin-conductivity into the sum of two terms, namely

\begin{equation}
\label{Full-Sigma}
\sigma_{xy}^{S,z}(\omega)=\sigma^{sH}_{c}(\omega)+\sigma^{sH}_{\tau}(\omega)
\end{equation}

\noindent in which the first one is given by

\begin{multline}
\label{Sigma-conv}
\sigma^{sH}_{c}(\omega)=\frac{\ e(\beta^2-\alpha^2)}{\ 4 \pi^2 m^*}\int_0^{2\pi}d\theta \frac{\ \cos^2\theta }{\ \Delta (\theta)}\int_{k^+_F (\theta)}^{k^-_F (\theta)} dk \\
\times
\frac{\ k^2}{\ (\omega+i\eta)^2-(2k\Delta(\theta)/\hbar)^2}
\end{multline}

%\begin{multline}
%\label{Sigma-conv}
%\sigma^{sH}_{c}(\omega)=\frac{\ e(\beta^2-\alpha^2)}{\ 4 \pi^2 %m^*}\int_0^{2\pi}d\theta \frac{\ \cos^2\theta }{\ \Delta (\theta)}\int_{k^+_F %(\theta)}^{k^-_F (\theta)} dk 
%\\ 
%       \times\frac{\ k^2}{\ (\omega+i\eta)^2-(2k\Delta(\theta)/\hbar)^2}
%\end{multline}

\noindent and comes from the conventional part of spin current definition. 
The second term is explicitly

%\begin{equation}\label{Sigma-tau}
%\sigma^{sH}_{\tau}(\omega)=\frac{\ e(\beta^2-\alpha^2)}{\ 4 \pi^2 m^*}\int_0^{2\pi}d\theta \frac{\ 8\cos^2\theta \Delta(\theta)}{\ \hbar^2}\int_{k^+_F (\theta)}^{k^-_F (\theta)}dk 
%\times 
%\frac{\ k^4 }{\ [(\omega+i\eta)^2-(2k\Delta(\theta)/\hbar)^2]^2}\Bigr]
%\end{equation}

\begin{multline}\label{Sigma-tau}
\sigma^{sH}_{\tau}(\omega)=\frac{\ e(\beta^2-\alpha^2)}{\ 4 \pi^2 m^*}\int_0^{2\pi}d\theta \frac{\ 8\cos^2\theta \Delta(\theta)}{\ \hbar^2}\int_{k^+_F (\theta)}^{k^-_F (\theta)}dk \\
        \times \frac{\ k^4 }{\ [(\omega+i\eta)^2-(2k\Delta(\theta)/\hbar)^2]^2}\Bigr]
\end{multline}

\noindent and arises from the spin torque contribution to the net 
spin-current. As a general trend, it will be shown later that such contribution introduces a significant change of the spin-Hall optical response overcoming the conventional part for typical sample parameters. 

At this stage it is worthwhile to remark that in the limit $\omega \tau  
\rightarrow \infty$ the impurity scattering does not play a significant 
role in spin-transport.\cite{Sinova1} Indeed, in the language of the 
diagrammatic technique, the resulting finite vertex corrections are 
negligible in the weak scattering limit for the  ac high frequency 
field regime. This occurs because in linear response to the ac field, 
the perturbative expansion in powers of $n_i$ (impurity density) of the 
spin-current-charge-current correlation function goes as $1/(\omega \tau)^n$ which makes  the contribution of the vertex corrections vanish at high frequencies.
\cite{Sinova1,Wang} Thus the linear response Kubo calculation without including explicitly the vertex corrections should give a qualitatively good agreement with the full diagrammatic technique at high enough frequencies and low impurity densities.\cite{Wang}

The $k-$integration in expressions \eqref{Sigma-conv} and 
\eqref{Sigma-tau} can be done analitycally, yielding for the 
conventional part of spin-conductivity 

%\begin{equation}
%\label{SigmaExact-conv}
%\frac{\ \sigma^{sH}_{c}(\omega)}{\ e/8\pi}=\frac{1}{\pi}\frac{\ \alpha^2-\beta^2}{\ \alpha^2+\beta^2}\int^{2\pi}_0 \frac{\cos^2(\theta)}{ \lambda(\theta)} \Bigr\{1
%+\frac{\ \hbar \tilde{\omega}}{\ 8 \varepsilon_{so} \lambda(\theta)} {\text {tanh}}^{-1}\Bigr[\frac{\ 8\varepsilon_{so}\lambda(\theta) \hbar \tilde{\omega}}{\ 16\varepsilon_F\varepsilon_{so}\lambda(\theta)-\hbar^2 \tilde{\omega}^2}\Bigr]\Bigr\}d\theta\\
%\end{equation}

\begin{multline}
\label{SigmaExact-conv}
\frac{\ \sigma^{sH}_{c}(\omega)}{\ e/8\pi}=\frac{1}{\pi}\frac{\ \alpha^2-\beta^2}{\ \alpha^2+\beta^2}\int^{2\pi}_0 \frac{\cos^2(\theta)}{ \lambda(\theta)} \Bigr\{1\\
+\frac{\ \hbar \tilde{\omega}}{\ 8 \varepsilon_{so} \lambda(\theta)} {\text {tanh}}^{-1}\Bigr[\frac{\ 8\varepsilon_{so}\lambda(\theta) \hbar \tilde{\omega}}{\ 16\varepsilon_F\varepsilon_{so}\lambda(\theta)-\hbar^2 \tilde{\omega}^2}\Bigr]\Bigr\}d\theta\\
\end{multline}

\noindent where $\tilde{\omega}=\omega+i\eta$ and $\lambda(\theta)$ as is defined in eq. \eqref{lambda}. The torque contribution becomes

%\begin{multline}
%\label{SigmaExact-tau}
% \frac{\ \sigma^{sH}_{\tau}(\omega)}{\ e/8\pi}=   -\frac{1}{\pi}\frac{\ \alpha^2-\beta^2}{\ \alpha^2+\beta^2}\int^{2\pi}_0 \frac{\cos^2(\theta)}{ \lambda(\theta)} \Bigr\{2
%     +\frac{\ 3\hbar\tilde{\omega}}{\ 8 \varepsilon_{so} \lambda(\theta)} {\text {tanh}}^{-1}\Bigr[\frac{\ 8\varepsilon_{so}\lambda(\theta) \hbar \tilde{\omega}}{\ 16\varepsilon_F\varepsilon_{so}\lambda(\theta)-\hbar^2\tilde{\omega}^2}\Bigr]\\
%     +\frac{\ [16\varepsilon_F\varepsilon_{so}\lambda(\theta)+\hbar^2\tilde{\omega}^2]\hbar^2\tilde{\omega}^2}{\ [16\varepsilon_F\varepsilon_{so}\lambda(\theta)-\hbar^2\tilde{\omega}^2]^2-64\varepsilon_{so}^2\lambda^2(\theta)\hbar^2\tilde{\omega}^2}\Bigr\}d\theta
%\end{multline}

\begin{multline}
\label{SigmaExact-tau}
 \frac{\ \sigma^{sH}_{\tau}(\omega)}{\ e/8\pi}=   -\frac{1}{\pi}\frac{\ \alpha^2-\beta^2}{\ \alpha^2+\beta^2}\int^{2\pi}_0 \frac{\cos^2(\theta)}{ \lambda(\theta)} \Bigr\{2\\
     +\frac{\ 3\hbar\tilde{\omega}}{\ 8 \varepsilon_{so} \lambda(\theta)} {\text {tanh}}^{-1}\Bigr[\frac{\ 8\varepsilon_{so}\lambda(\theta) \hbar \tilde{\omega}}{\ 16\varepsilon_F\varepsilon_{so}\lambda(\theta)-\hbar^2\tilde{\omega}^2}\Bigr]\\
     +\frac{\ [16\varepsilon_F\varepsilon_{so}\lambda(\theta)+\hbar^2\tilde{\omega}^2]\hbar^2\tilde{\omega}^2}{\ [16\varepsilon_F\varepsilon_{so}\lambda(\theta)-\hbar^2\tilde{\omega}^2]^2-64\varepsilon_{so}^2\lambda^2(\theta)\hbar^2\tilde{\omega}^2}\Bigr\}d\theta
\end{multline}

\noindent Note that if $\beta=0$ (or $\alpha=0$), {\it i.e.} if
$\lambda(\theta)=1$, the angular dependence of the integrand 
above reduces significantly and leads to a close analytic form for 
$\sigma^{sH}_{c}(\omega)$ and $\sigma^{sH}_{\tau}(\omega)$. For the case 
$\alpha\ne\beta$ the $\theta-$integrals cannot be performed 
straightforwardly  and a numerical integration has to be implemented. 
Thus, it is convenient to consider first the physically reasonable limit $k_{so}\ll k_F$, which typically holds for 2DEGs in III-V based semiconductor heterostructures, and for which an analytical expression for the frequency dependent spin-conductivity can be derived.

From \eqref{Sigma-conv} and \eqref{Sigma-tau}, we obtain to leading
order in $k_{so}/k_F$,

\begin{equation}\label{analiticatodo}
\sigma_{xy}^{S,z}(\omega)\simeq\sigma_{c,o}^{sH}(\omega)+\sigma_{\tau,o}^{sH}(\omega),
\end{equation}

\noindent in which the conventional part reads

\begin{equation}\label{analitica-conv}
\frac{\ \sigma_{c,o}^{sH}(\omega)}{\ e/8\pi}=-\frac{\ 16 (\beta^2-\alpha^2)}{\ (\beta^2+\alpha^2)} \frac{\ \varepsilon_F \varepsilon_{so}}{\ \prod_{\mu} [{\xi^2_{\mu}-\hbar^2\tilde{\omega}^2}]^{1/2}}\, ,
\end{equation}

\noindent while the spin-torque part takes the form

%\begin{equation}\label{analitica-tau}
%\frac{\ \sigma_{\tau,o}^{sH}(\omega)}{\ e/8\pi}=\frac{\ 32(\beta^2-\alpha^2)}{\ (\beta^2+\alpha^2)}\Bigr[ \frac{\ \varepsilon_F \varepsilon_{so}}{\ \prod_{\mu} [{\xi^2_{\mu}-\hbar^2\tilde{\omega}^2}]^{1/2}}
%+\frac{\ \varepsilon_F \varepsilon_{so} \hbar^2 \tilde{\omega}^2[16\varepsilon_F \varepsilon_{so}-\hbar^2\tilde{\omega}^2] }{\ \prod_{\mu} [{\xi^2_{\mu}-\hbar^2\tilde{\omega}^2}]^{3/2}}\Bigr],
%\end{equation}

\begin{multline}\label{analitica-tau}
\frac{\ \sigma_{\tau,o}^{sH}(\omega)}{\ e/8\pi}=\frac{\ 32(\beta^2-\alpha^2)}{\ (\beta^2+\alpha^2)}\Bigr[ \frac{\ \varepsilon_F \varepsilon_{so}}{\ \prod_{\mu} [{\xi^2_{\mu}-\hbar^2\tilde{\omega}^2}]^{1/2}}\\
+\frac{\ \varepsilon_F \varepsilon_{so} \hbar^2 \tilde{\omega}^2[16\varepsilon_F \varepsilon_{so}-\hbar^2\tilde{\omega}^2] }{\ \prod_{\mu} [{\xi^2_{\mu}-\hbar^2\tilde{\omega}^2}]^{3/2}}\Bigr],
\end{multline}

\noindent with $\xi_\mu = 2k_F\mid \alpha+\mu\beta \mid $.

Our result \eqref{analitica-conv} agrees with  
eq. (39) of Ref. [\onlinecite{Erlingsson}] obtained there via 
spin-susceptiblity calculation.\cite{Catalina-etal} 
Note also that the second term of \eqref{analitica-tau} vanishes
for $\tilde{\omega}=0$, while the first one, when added to
\eqref{analitica-conv}, reverses the sign of the static value of the spin Hall
conductivity \eqref{analiticatodo}. On the other hand,
it will be seen that, at finite frequencies, the torque dipole 
contribution of Shi {\it et al.}\cite{Shi} generally dominates 
over the conventional spin current contribution.

It is illustrative to analyze the behavior of the spin-Hall conductivity
in the static limit ($\omega=0$) in presence of weak disorder when the 
new conserved spin-current operator is applied. 
In particular, from eqs. \eqref{analitica-conv} and \eqref{analitica-tau}, 
the static value, for the pure Rashba coupling case, can be written in the appealing form as

\begin{equation}\label{del dyakonov}
\frac{\ \sigma_{xy}^{S,z}(0)}{\ e/8\pi}=\frac{\ \eta/\eta_{so}-1}
{\ (\eta/\eta_{so}+1)^2}
\end{equation}

\noindent where $\eta_{so}=\tau_{so}^{-1}$, being $\tau_{so}$ the  
Dyakonov-Perel spin-orbit relaxation time \cite{Dyakonov} with  
$\tau^{-1}_{so}=\frac{\ (2\alpha k_F/\hbar)^2}{\tau^{-1}}$.  
Since  $\eta$ is typically smaller than $\eta_{so}$ 
($i.e.$ $2\alpha k_F>\hbar\eta$) note that the static value of 
$\sigma_{xy}^{S,z}$ is always negative. 
 
It is known however that the dc limit is problematic within the Kubo 
formula when using the conventional spin-current operator, namely because it 
leads to the incorrect physics for $\varepsilon_{so}\gg\hbar\eta$ as a 
result of neglecting the contribution of the vertex corrections. 
This is not necessarily the case  for finite frequencies and 
relatively low impurity densities (as discussed below), 
which is in fact the regime that we are mostly interested in here.
In the opposite limit, $\varepsilon_{so}\ll\hbar\eta $,
Kubo formula gives a result which coincides qualitatively with
the expected result [\onlinecite{Schliemann}]. 
We first consider this case. 
Taking the limit $\omega\rightarrow 0$ for finite but weak scattering 
mechanism with a time-relaxation rate $\tau^{-1}=\eta $ and   
expanding \eqref{analitica-conv} and \eqref{analitica-tau} 
in powers of $\varepsilon_{so}/\hbar \eta\ll 1$ to lowest order we get

\begin{equation}
\frac{\sigma _{c,o}^{sH}(0)}{\ e/8\pi}
\simeq  \frac{  \alpha ^{2}-\beta ^{2}}{\alpha ^{2}+\beta ^{2}}
\left( \frac{16\varepsilon _{F}\varepsilon_{so}}{\hbar ^{2}\eta ^{2}}\right) 
\end{equation}

\noindent being $\sigma _{\tau,o}^{sH}(0)=0$ 
to first order. 

Now, if the impurity scattering is weak compared to 
spin-orbit coupling, $\varepsilon_{so}/\hbar\eta\gg 1$, we obtain  
to first order,

\begin{equation}
\frac{\sigma^{sH}_{c,o}(0)}{\ e/8\pi} \simeq 
\left [ 1  
-\frac{\ \hbar^2\eta^2 }{\ 16\varepsilon_F\varepsilon_{so}} 
\frac{\ (\alpha^2+\beta^2)^2}{\ (\alpha^2-\beta^2)^2}
 \right ] \, {\text {sign}}(\alpha^2-\beta^2)\\
\end{equation}

\noindent with a nonzero torque part 

\begin{equation}
\frac{\sigma^{sH}_{\tau,o}(0)}{\ e/8\pi} \simeq 
\left [ -2  
+\frac{\ \hbar^2\eta^2 }{\ 4\varepsilon_F\varepsilon_{so}} 
\frac{\ (\alpha^2+\beta^2)^2}{\ (\alpha^2-\beta^2)^2}
 \right ]\, {\text {sign}}(\alpha^2-\beta^2)\\
\end{equation}

The above expressions for $\sigma^{sH}_{c,o}(0)$
reduce, in each case, to the known formulas for $\beta=0$ reported in Ref.
[\onlinecite{Schliemann}].

\section{RESULTS AND DISCUSSIONS}

We first study the case where only the Rashba coupling is present 
($\beta=0$). In Fig. 2 we plot the real part of the spin-Hall 
conductivity versus the frequency of the applied electric field
as obtained from eqs. \eqref{SigmaExact-conv} and 
\eqref{SigmaExact-tau} ($\lambda(\theta)=1$).
The Rashba parameter has been fixed to $\alpha=1.6\times 10^{-9}$ eVcm 
which is a typical experimental value for a 2DEG in InAs-based quantum wells. 
Using an electron effective mass of $m^*=0.055\,m_o$, these values give us a 
characteristic Rashba energy of $\varepsilon_{so}=0.092$ meV with a 
spin-splitting at the Fermi energy of $\Delta_{R}=2\alpha k_F \simeq 5.6$ meV. 
The corresponding Fermi wave number is estimated with $k_F=\sqrt{2\pi n}$ 
considering a moderated sheet electron density of 
$n=5\times 10^{11}$cm$^{-2}$. The parameter describing the momentum 
relaxation rate has been chosen such that $\hbar \eta =0.4$ meV, value that 
corresponds to high quality samples with mobilities 
$\mu= e \tau/m^*\simeq 5$ m$^2$/Vs and relaxation times of 
$\tau\simeq 1.6$ ps. We have plotted as a reference 
(dotted curve in Fig. 1) the result obtained using the conventional 
definition of spin-current [eq. \eqref{Sigma-conv}]. 

The spin Hall conductivity 
$\sigma_{c}^{sH}(\omega)$ shows resonance behavior at energies
$\hbar\omega_+\simeq 2\alpha k_F^+$ and $\hbar\omega_-\simeq 2\alpha k_F^-$,
which correspond, respectively,
 to the minimum and maximum photon energy required
to induce optical transitions between the initial $\mu=-1$ and
final $\mu=+1$ spin-splitted branches (see. Fig.1).
At low frequencies it approaches the universal value $e/8\pi$, while
it vanishes for high frequencies.
Because of the finite value of the damping parameter $\eta$,
the spectrum has a smoothed shape across the shifted resonance
frequencies $\omega_{\pm}'=\sqrt{\omega_{\pm}^2+\eta^2}$.
%\cite{approximate-omegas} 
Notice that for $k_{so}\ll k_F$ the approximated  solution of $\sigma_{c,o}^{sH}(\omega)$ in Eq. \eqref{analitica-conv} will give nearly the same numerical values with sligtly shifted resonance frequencies at $\omega_{\pm}'\simeq\sqrt{\omega_{o}^2 \pm 2 \eta^2 \sqrt{(\Delta_R/\hbar \eta)^2+1}}$, where $\eta>0$ is understood and   $\omega_o=\eta\sqrt{(\Delta_R/\hbar \eta )^2+1}$ is the intermediate frequency  at which the conventional part vanishes.

For the sake of contrast, the inset of Fig. 1 
shows the behavior of the conventional part of the spin-Hall conductivity 
(eq.\eqref{Sigma-conv}) and that obtained from the expression 
(21) of Ref. [\onlinecite{Chalaev-Loss}] which incorporates explicitly 
the vertex corrections. 
It is shown that at finite high frequencies 
($\hbar\omega\gtrsim 0.5$ meV) the result obtained neglecting the 
vertex corrections reproduce qualitatively the expected behavior, 
being in quite good agreement for finite frequencies outside the 
corresponding energy window referred above, that is for 
$\omega \gtrsim \omega'_{-}$and $\omega \lesssim \omega'_{+}$

As for the effect of the torque dipole contribution on the
spin Hall conductivity we notice  
that the first two terms in the expression \eqref{SigmaExact-tau}
change the sign of the terms giving $\sigma_{c}^{sH}(\omega)$
in \eqref{SigmaExact-conv}, when added to obtain the total
conductivity $\sigma_{xy}^{S,z}(\omega)$. On the other hand, 
the last term of $\sigma_{\tau}^{sH}(\omega)$ (eq.\eqref{SigmaExact-tau})
is a rational function which can be rewritten (for $\beta=0$ and
$\tilde{\omega}\neq 0$) as
\begin{equation}
-\frac{(\tilde{\omega}^2+\Delta_R^2/\hbar^2)
(\tilde{\omega}^2+4\varepsilon_R^2/\hbar^2)}
{(\tilde{\omega}^2-\omega^2_+)(\tilde{\omega}^2-\omega^2_-)} \ \ ,
\end{equation}  
where $\varepsilon_R=2\varepsilon_{so}=m^*\alpha^2/\hbar^2$.
As is shown in Fig.2, this contribution dominates the shape of the
spectrum. It also resonates at the (shifted) frequencies $\omega_{\pm}'$,
being positive for $\omega$ between $\omega_+'$ and $\omega_-'$,
and negative otherwise. Consequently, a dramatic change 
of the overall shape of the spectrum is observed.
This is one of the main results of this paper.  By measuring if possible, 
the spin Hall current/accumulation at low temperatures in the frequency domain, it could be 
helpful perhaps, to establish the validity of the definition of the spin 
current operator proposed by J. Shi {\it et al.} \cite{Shi} by contrasting with our results.

\begin{figure} 
\centerline{\includegraphics[width=3 in]{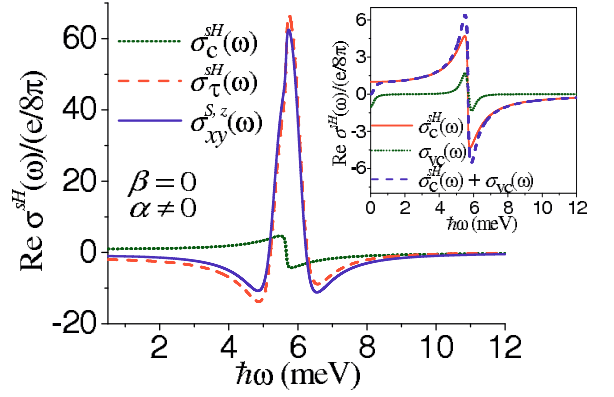}}
\caption{(Color online) Real part of the spin Hall conductivity as a 
function of the photon energy for a 2DEG system with Rashba SOI only ($\beta=0$). The dotted curve is obtained using the conventional definition of 
spin-current, whereas the dashed and solid curves are the torque 
contribution and 
the total spin conductivity, respectively, employing the conserved 
spin-current operator. In the inset the result obtained taking into 
account vertex corrections is contrasted with our calculations. See text 
for parameter values used. }
\label{Fig2}
\end{figure}

\begin{figure}  
\centerline{\includegraphics[width=2.6 in]{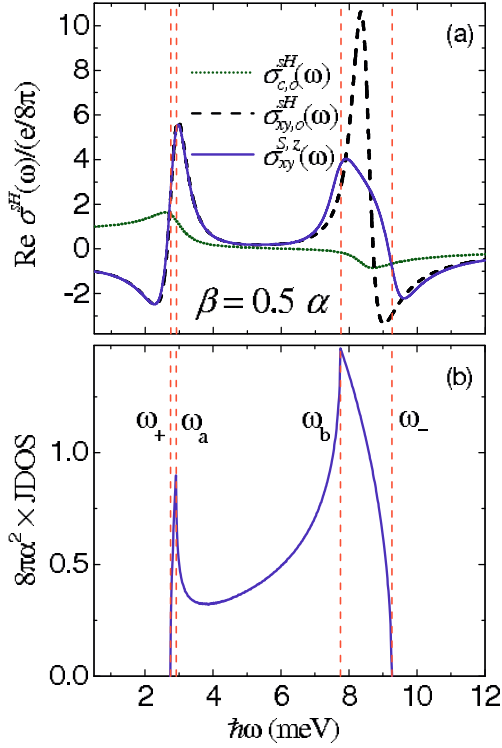}}
\caption{(Color online)  (a) Frequency dependent spin Hall conductivity 
for a 2DEG system  with finite Rashba and Dresselhaus SOI 
($\beta = 0.5\alpha$). Employing, the conventional definition of 
spin-current (dotted curve),  the new conserved spin-current 
operator ${\cal J}^{S,z}_{xy}$ in the limit $k_{so}\ll k_F$ (dashed curve) 
and by the exact numerical integration of eq. \eqref{kubo} (solid curve), 
respectively. Here $\hbar\eta=0.25$ meV and other parameters  as in Fig. 2.  
(b) Plot for the corresponding joint density of states (JDOS). There are four main   frequencies here, two   defining the optical absorption edges, $\omega_{-}$ and $\omega_{+}$, while the other two correspond to the peaks of the JDOS 
occuring at $\omega_{a}$ and $\omega_{b}$. The latter two arises due to the symmetry of the spin-split conduction bands in $k-$space at the Fermi level. The dashed vertical lines at the frequencies referred above are here to guide eye. }
\label{Fig3}
\end{figure}

New interesting features appear when the interplay of the Rashba and Dresselhaus SOI is considered. In Fig. 3(a) we plot the frequency dependent spin-Hall conductivity for  $\beta=0.5\alpha$, here 
$\hbar \eta = 0.25$ meV and the remaining parameters are as in Fig. 2. The results using the approximated formulas 
\eqref{analitica-conv} and \eqref{analitica-tau} and the 
exact numerical integration of \eqref{SigmaExact-conv} and 
\eqref{SigmaExact-tau} are presented for comparison. Here is also evident 
the remarkable difference between the optical spectrum resulting from the 
use of the standard and the new conserved spin-current operator definition, ${\cal J}_x^{S,z}$. 
In addition, the spectra become wider and highly asymmetric
in comparison with those of the $\beta=0$ or $\alpha=0$ case. 
As was recently discussed in 
Ref.\,\onlinecite{Maytorena-etal}, the main spectral features can
be understood as due to the anisotropic spin splitting caused by the
simultaneous presence of the Rashba and Dresselhaus couplings.
In the limit of vanishing temperature, the sum over states 
in eq. \eqref{kubo} is restricted to the region between the
Fermi contours $k_F^+(\theta)\leq k\leq k_F^-(\theta)$, for which
$\varepsilon_-(k,\theta)\leq\varepsilon_F\leq\varepsilon_+(k,\theta)$,
(see Fig.1). Certain distinctive frequencies can be identified 
given the anisotropic $k-$space available for the optical response.
To illustrate this, we consider the joint density of
states (JDOS) for the spin-split branches (Fig.3b), which gives the number
of direct transitions that can take place at the energy $\hbar\omega$.
These transitions involve only states with wave vectors that satisfy
the equation $\varepsilon_+(k,\theta)-\varepsilon_-(k,\theta)-\hbar\omega=0$,
which for a fixed photon energy, defines a curve $C_r(\omega)$ in 
$k-$space describing a rotated ellipse with semi axis of lengths
$k_a(\omega)=\hbar\omega/2|\alpha-\beta|$ and 
$k_b(\omega)=\hbar\omega/2|\alpha+\beta|$ oriented along the
principal axes (1,1) and (-1,1), respectively (see Fig.1).
Thus, for our problem, the JDOS involves states only along the
arcs of the resonance curve $C_r(\omega)$ lying within the mentioned region
$k_F^+(\theta)\leq k\leq k_F^-(\theta)$. 
The peaks observed in the JDOS correspond to energy transitions involving
states in the vicinity of the symmetry points
$k_a(\omega)=k_F^-(\pi/4)$ and $k_b(\omega)=k_F^+(3\pi/4)$,
for which the energy splitting reaches extreme values.
These equations determine two energies 
$\hbar\omega_a=2k_F^-(\pi/4)|\alpha-\beta|=
2k_F|\alpha-\beta|+2m^*(\alpha-\beta)^2/\hbar^2$
and $\hbar\omega_b=2k_F^+(3\pi/4)|\alpha+\beta|=
2k_F|\alpha+\beta|-2m^*(\alpha+\beta)^2/\hbar^2$. 
Similarly, we can see that there are absorption edges
at energies $\hbar\omega_{\pm}=2k_F|\alpha\mp \beta|
\mp 2m^*(\alpha\mp \beta)^2/\hbar^2$,
corresponding to transitions between states at the points
$k_a(\omega)=k_F^+(\pi/4)$ and $k_b(\omega)=k_F^-(3\pi/4)$. For clarity
such energies are indicated in Fig.3 with dashed vertical lines.
As expected, the spin Hall conductivity $\sigma_{xy}^{S,z}(\omega)$
also shows structure at the photon energies $\hbar\omega_{\pm}$,
$\hbar\omega_a$ and $\hbar\omega_b$. The finite value
of $\hbar \eta$ chosen here yields a slight shifting of these energies and
an overall smoothing of the spectrum.

\begin{figure} 
\centerline{\includegraphics[width=2.6 in]{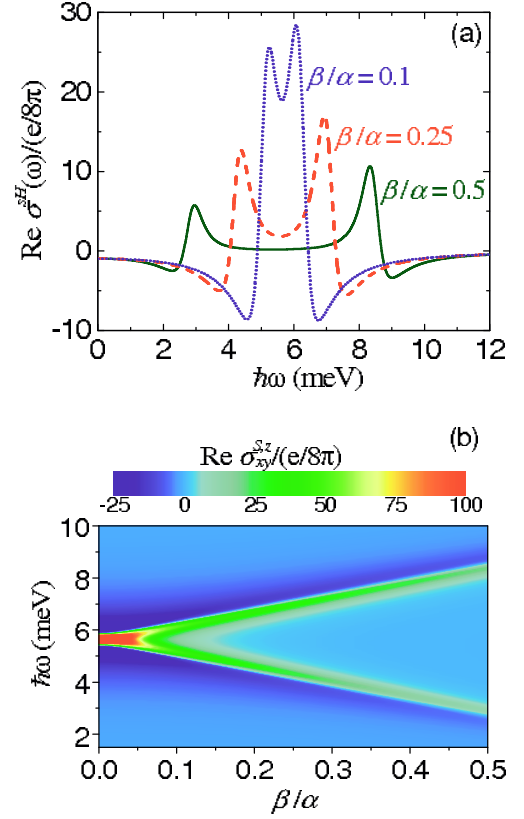}}
\caption{(Color online) (a) Spin Hall conductivity for 
$\beta/\alpha = 0.5,0.25,0.1$ (solid, dashed and dotted curves, 
respectively) with the proposed conserved spin-current operator 
${\cal J}^{S,z}_{xy}$.   The parameters are as in Fig. 2 with $\hbar\eta=0.25$ 
meV. Notice that the resonance peaks tend to separate in energy as the 
ratio $\beta/\alpha$ is increased while its magnitude is diminished. 
(b) Color contour map of the ac spin-Hall conductivity showing its behavior with a continuous variation of the frequency $\omega$ and to the ratio $\beta/\alpha$. }
\label{Fig4}
\end{figure}

It is clear that the new definition of spin-current operator 
by Shi {\it et al.} \cite{Shi} yields dramatically different frequency 
response from that predicted by the conventional definition. 
In addition,  as it occurs with
the pure Rashba (or Dresselhaus) SOI case, the torque dipole contribution 
(not shown) turns out to be the dominant term in the spin-Hall 
conductivity. Notice that the spectra in Fig.3(a) show that 
the magnitude and the direction of the dynamic spin Hall current 
strongly depends on the frequency  and on the spin-orbit coupling 
strengths $\alpha$ and $\beta$, suggesting thus its  control  via electrical 
gating (by varying the Rashba coupling) and/or by adjusting the electric 
field frequency.

We have also explored the effect induced of varying the ratio $\beta/\alpha$  on the conserved  spin-Hall conductivity as a function of the exciting frequency. In particular, in Fig. 4(a) the spin-Hall 
response is shown for the specific values of $\beta/\alpha = 0.5,0.25$ 
and $0.1$ while fixing the rest of the parameters as in Fig.3. Notice that the energy separation of the  resonance peaks becomes larger and the intensity of the peaks get diminished as the aspect ratio $\beta/\alpha$ is increased. Such effect is emphasized in panel 4(b) where  a color map of the spin-Hall conductivity is plotted as a function of a continuous variation of the ratio $\beta/\alpha$ and $\omega$.   Such behavior responds to the widespreading ($\Delta{\cal E}=\hbar\omega_{-}-\hbar\omega_{+}$) of absorption bandwidth  with $\beta$. \cite{Maytorena-etal}

\section{SUMMARY}

In summary, we have examined the spin-Hall conductivity in the frequency regime for a two dimensional electron gas with Rashba and Dresselhaus spin-orbit interaction employing a recently proposed form for a conserved spin current operator.  Our results shows that the optical spectrum of the spin Hall conductivity changes substantially when the new conserved spin current operator is used.  It is predicted that the torque dipole contribution typically overcomes the conventional part of the total spin Hall conductivity.  In addition, it is shown that the magnitude and the direction of the dynamic spin Hall current is rather sensitive to the frequency and to  the spin-orbit ($\alpha$ and $\beta$) coupling strengths due to the spin torque contribution. We hope that these results will encourage experimentalists to measure the spin Hall accumulation and/or spin density currents and to explore to what extent the new definition of the spin current operator provides a satisfactory description of the ac spin Hall conductivity in such systems.

\section{acknowlegements}F.M. is thankful to Q. Niu for useful comments. This 
work was supported by CONACyT-Mexico grant J41113F, and by  DGAPA-UNAM IN113807-3.

\appendix 
\section{}
 
Here we briefly outline the derivation of the expectation value for spin-current-charge-current correlation function  $\langle \psi_{{\bf k},\mu}
(\textbf{r}) \vert [{\cal J}_x^{S,z}(t), v_y (0)] \vert \psi_{{\bf k},\mu}
(\textbf{r}) \rangle\equiv{\cal F}_{\mu}({\bf k},t)$. We begin by writing the spin current operator at all times in the Heisenberg picture, ${\cal J}_x^{S,z}(t)= e^{iHt/ \hbar}{\cal J}_x^{S,z}(0)e^{-iHt/ \hbar}$,  with ${\cal J}_x^{S,z}(0)$ as given by \eqref{corrienteespin}. Similarly is done for the operators ${x}(t)$, $\sigma_x (t)$ and $\sigma_y (t)$, which after some algebraic manipulations explicitly reads,

\begin{equation}
\begin{split}
\quad\quad {x}(t)&={x}(0)+t\Bigr(\frac{\ p_x}{\ m}+\frac{\ \alpha \sigma_y+\beta \sigma_x}{\ \hbar}\Bigr)\\
          & \quad +\frac{\ \hbar}{\ 2 
\Delta^2p^2}(\beta^2-\alpha^2)p_y\sigma_z\Bigr[\cos\Big(\frac{\ 2\Delta p}{\ \hbar^2} t\Bigr)-1\Bigr] \\
          & \quad - \frac{\ \hbar}{\ 2\Delta^3 p^3} (\beta^2-\alpha^2)p_y \Bigr[\frac{\ 2t\Delta p}{\ \hbar^2}-\sin\Bigr(\frac{\ 2\Delta p}{\ \hbar^2} t\Bigr) \Bigr]{\cal K}\, ,
\end{split}
\end{equation}

\begin{equation}
\begin{split}
\sigma_x(t)&=\sigma_x(0)+\frac{\ (\alpha p_x-\beta p_y)}{\ \Delta p}\sigma_z\sin\Bigr(\frac{\ 2\Delta p}{\ \hbar^2} t\Bigr)\\
          & \quad -\frac{\ (\alpha p_x-\beta p_y)}{\ \Delta^2 p^2}\Bigr[\cos\Bigr(\frac{\ 2\Delta p}{\ \hbar^2}t\Bigr)-1\Bigr]{\cal K}\, ,
\end{split}
\end{equation}

\noindent and 

\begin{equation}
\begin{split}
\sigma_y(t)&=\sigma_y(0)-\frac{\ (\beta p_x-\alpha p_y)}{\ \Delta p}\sigma_z\sin\Bigr(\frac{\ 2\Delta p}{\ \hbar^2} t\Bigr) \\
          & \quad -\frac{\ (\beta p_x-\alpha p_y)}{\ \Delta^2 p^2} \Bigr[\cos\Bigr(\frac{\ 2\Delta p}{\ \hbar^2}t\Bigr)-1\Bigr]{\cal K}
\end{split}
\end{equation}

\noindent with  ${\cal K}=\alpha(\sigma_x p_x+\sigma_y p_y)-\beta(\sigma_x p_y+\sigma_y p_x)$, and in which  $\vec \sigma$ and $\vec p$ are given in the Schr\"odinger picture. The expressions above are needed in the calculation of the commutator $[{\cal J}_x^{S,z}(t),v_y(0)]$ which gives

%\begin{equation}
%\begin{split}
%[{\cal J}_x^{S,z}(t),v_y(0)] & =\frac{\ (\alpha \sigma_x+\beta \sigma_y)}{\ 2m^*} e^{iHt/\hbar}\sigma_z p_x e^{-iHt/\hbar} 
% -  \,e^{iHt/\hbar}\sigma_z p_x e^{-iHt/\hbar}\frac{\ (\alpha \sigma_x+\beta \sigma_y)}{\ 2m^*}\\
%& +i\Bigr\{x(0), {\cal L}\Bigr\}+\frac{\ 2itp_x}{ m^*}{\cal L}
%\end{split}
%\end{equation}

\begin{multline}
[{\cal J}_x^{S,z}(t),v_y(0)]=\frac{\ (\alpha \sigma_x+\beta \sigma_y)}{\ 2m^*} e^{iHt/\hbar}\sigma_z p_x e^{-iHt/\hbar} \\
\quad -  \,e^{iHt/\hbar}\sigma_z p_x e^{-iHt/\hbar}\frac{\ (\alpha \sigma_x+\beta \sigma_y)}{\ 2m^*}+i\Bigr\{x(0), {\cal L}\Bigr\}+\frac{\ 2itp_x}{ m^*}{\cal L}
\end{multline}

\noindent with ${\cal L}=\frac{\ \Delta p}{\ \hbar^2}(\alpha\sigma_y-\beta\sigma_x)\sin\Bigr(\frac{\ 2\Delta p}{\ \hbar^2}t \Bigr)+\frac{\ \sigma_z}{\ \hbar^2}{\cal P}\cos\Bigr(\frac{\ 2\Delta p}{\ \hbar^2}t\Bigr )$ and ${\cal P}= 2\alpha\beta p_x-(\alpha^2+\beta^2)p_y$. Calculating the expectation value of (A4) using \eqref{eigenvectors} we finally  arrive to ${\cal F}_{\mu}({\bf k},t)$ as defined in equation \eqref{valoresperado}.

% Set the ending of a LaTeX document
\end{document}